\documentclass{article}
\usepackage{graphicx}
\usepackage{epsfig}
\usepackage{url}
\usepackage{psfrag}
\usepackage{times}
\usepackage{amsfonts,amssymb,amsbsy}
\usepackage{amsmath}
\usepackage{cite}
\usepackage{latexsym}
\usepackage{amsfonts,amsbsy}
\usepackage{xcolor} 
\usepackage{tikz}

\begin{document}

\title{Self-Dual Boundary Conditions\\ in Electromagnetics}
\author{Ismo V. Lindell and Ari~Sihvola}%
        \date{School of Electrical Engineering,\\ Aalto University, Espoo, Finland\\ 
{\tt ismo.lindell@aalto.fi}\\ {\tt ari.sihvola@aalto.fi} }
\pagestyle{myheadings}



\def\e{\begin{equation}} 
\def\f{\end{equation}} 
\def\ea{\begin{eqnarray}} 
\def\fa{\end{eqnarray}} 

\def\##1{{\mbox{\textbf{#1}}}}
\def\%#1{{\mbox{\boldmath $#1$}}}
\def\=#1{{\overline{\overline{\mathsf #1}}}}
\def\RR{\mbox{\boldmath $\R$}}
\def\nn#1{{\sf #1}}
\def\SE{{\mathbb E}}
\def\SF{{\mathbb F}}

\def\*{^{\displaystyle*}}
\def\xx{\displaystyle{{}^\times}\llap{${}_\times$}}
\def\.{\cdot}
\def\x{\times}
\def\oo{\infty}

\def\D{\nabla}
\def\d{\partial}

\def\ra{\rightarrow}
\def\lra{\leftrightarrow}
\def\Ra{\Rightarrow}
\def\le{\left(}
\def\ri{\right)}
\def\l#1{\label{eq:#1}}
\def\r#1{(\ref{eq:#1})}
\def\am{\left(\begin{array}{c}}
\def\amm{\left(\begin{array}{cc}}
\def\ammm{\left(\begin{array}{ccc}}
\def\ammmm{\left(\begin{array}{cccc}}
\def\a{\end{array}\right)}

\def\I{\int\limits}
\def\OI{\oint\limits}

\def\A{\alpha}
\def\B{\beta}
\def\de{\delta}
\def\De{\Delta}
\def\E{\epsilon}
\def\g{\gamma}
\def\G{\Gamma}
\def\h{\eta}
\def\K{\kappa}
\def\la{\lambda}
\def\La{\Lambda}
\def\M{\mu}
\def\o{\omega}
\def\Om{\Omega}
\def\R{\rho}
\def\s{\sigma}
\def\t{\tau}
\def\z{\zeta}
\def\X{\chi}
\def\TH{\theta}
\def\Th{\Theta}
\def\VF{\varphi}
\def\VR{\varrho}
\def\VT{\vartheta}
\def\ve{\%\varepsilon}

\def\tr{{\rm tr }}
\def\spm{{\rm spm}}
\def\det{{\rm det}}
\def\Det{{\rm Det}}
\def\sgn{{\rm sgn}}
\def\bi{\bibitem}

\def\W{\wedge}
\def\WW{\displaystyle{{}^\wedge}\llap{${}_\wedge$}}
\def\Adj{{\rm Adj\mit}}
\def\ua{\uparrow}
\def\da{\downarrow}
\def\uda{\updownarrow}

\def\J{\rfloor}
\def\L{\lfloor}
\def\JJ{\rfloor\rfloor}
\def\LL{\lfloor\lfloor}

\maketitle\

\begin{abstract}
Invariance in duality transformation, the self-dual property, has important applications in electromagnetic engineering. In the present paper, the problem of most general linear and local boundary conditions with self-dual property is studied. Expressing the boundary conditions in terms of a generalized impedance dyadic, the self-dual boundaries fall in two sets depending on symmetry or antisymmetry of the impedance dyadic. Previously known cases are found to appear as special cases of the general theory. Plane-wave reflection from boundaries defined by each of the two cases of self-dual conditions are analyzed and waves matched to the corresponding boundaries are determined. As a numerical example, reflection from a special case, the self-dual EH boundary, is computed for two planes of incidence. 
\end{abstract}

\section{General Form of Boundary Conditions}

In \cite{PIERL2016,AP2017,GBC} the set of most general linear and local boundary conditions (GBC) was introduced as
\e \amm \#a_1 & \#b_1\\ \#a_2 & \#b_2\a \.\am \#E\\ \h_o\#H\a = \am 0\\0\a, \l{GBC} \f
where $\#a_1, \#a_2, \#b_1$ and $\#b_2$ are four dimensionless vectors and $\h_o=\sqrt{\M_o/\E_o}$. It is assumed for simplicity that the medium outside the boundary is isotropic with parameters $\E_o,\M_o$. 

The general form of conditions \r{GBC} were arrived at through a process of generalizing known boundary conditions. Denoting vector tangential to the boundary surface by $()_t$, the conditions of perfect electric (PEC) and magnetic (PMC) conductor conditions are respectively defined by
\e \#a_{1t}\.\#E=0,\ \ \ \ \#a_{2t}\.\#E=0,\ \ \ \ \#a_{1t}\x\#a_{2t}\not=0,\l{PEC} \f
and
\e \#b_{1t}\.\#H=0,\ \ \ \ \#b_{2t}\.\#H=0,\ \ \ \ \#b_{1t}\x\#b_{2t}\not=0. \l{PMC}\f
A generalization of these, the perfect electromagnetic conductor (PEMC), is defined by \cite{PEMC,253}
\e \#b_{1t}\.(\#H + M\#E)=0,\ \ \ \#b_{2t}\.(\#H+ M\#E)=0, \l{PEMC}\f
where $M$ is the PEMC admittance.

Conditions \r{PEC} -- \r{PEMC} are special cases of the impedance-boundary conditions \cite{Methods} defined by,
\e  \amm \#a_{1t} & \#b_{1t}\\ \#a_{2t}& \#b_{2t}\a \.\am \#E\\ \h_o\#H\a = \am 0\\ 0\a. \l{imp} \f
The soft-and-hard (SH) boundary \cite{SH,SH1}
\e \#a_t\.\#E=0,\ \ \ \ \#a_t\.\#H=0, \l{SH}\f
and the generalized soft-and-hard (GSH) boundary  \cite{GSH},
\e \#a_t\.\#E=0,\ \ \ \ \#b_t\.\#H=0, \f
are other special cases of the impedance boundary. 

As examples of boundaries not special cases of \r{imp}, the DB boundary is defined by \cite{Rumsey,DB,259}
\e \#n\.\#D= \E_o\#n\.\#E=0,\ \ \ \ \#n\.\#B = \M_o\#n\.\#H=0, \l{DB}\f
while the soft-and-hard/DB (SHDB) boundary \cite{SHDB} generalizes both the SH and the DB boundaries as
\e \amm \A\#n & \#a_t\\-\#a_t & \A\#n\a \.\am \#E\\ \h_o\#H\a = \am 0 \\ 0\a. \l{SHDB}\f
A further generalization is the generalized  soft-and-hard/DB (GSHDB) boundary \cite{GSHDB}, with
\e \amm a_{1n}\#n & \#b_{1t}\\ \#a_{2t}&  b_{2n}\#n\a\.\am \#E\\ \h_o\#H\a=\am 0\\ 0\a. \l{GSHDB}\f

A remarkable property of the GSHDB boundary and its special cases is that any plane wave can be decomposed in two parts reflecting from the GSHDB boundary as from respective PEC and PMC boundaries \cite{GSHDB}. The converse was shown in \cite{PIERL2016}, i.e., that a GBC boundary, required to have this property, must actually equal a GSHDB boundary. 

The form \r{GBC} of general boundary conditions is not unique, since the same boundary is defined by conditions  obtained by multiplying the vector matrix by any scalar matrix, 
\e \amm \#a_1 & \#b_1\\ \#a_2 & \#b_2\a  \ra \amm \A & \B \\ \g & \de\a \amm \#a_1 & \#b_1\\ \#a_2 & \#b_2\a , \f
with nonzero determinant $\A\de-\B\g\not=0$. A more unique form of general boundary conditions \r{GBC} can be written as
\e \#m\x\#E = \=W\.\h_o\#H. \l{WH} \f
In fact, \r{WH} is equivalent to \r{GBC} for
\ea \#m &=& \#a_1\x\#a_2, \l{m}\\
 \=W &=& \#a_1\#b_2-\#a_2\#b_1. \l{W}\fa
Here we must assume $\#m=\#a_1\x\#a_2\not=0$, which rules out a special case of \r{GBC}. The conditions of a boundary with $\#a_1\x\#a_2=0$ in \r{GBC} can be reduced to the form 
\e \amm \#a & \#b_1\\ 0 & \#b\a \am \#E\\ \h_o\#H\a = \am 0\\ 0\a, \l{m=0} \f
in terms of three vectors $\#a,\#b$ and $\#b_1$, see the Appendix. For the special case $\#b_1=0$, \r{m=0} is reduced to
\e \amm \#a & 0\\ 0 & \#b\a \am \#E\\ \h_o\#H\a = \am 0\\ 0\a, \l{EH} \f
corresponding to what have been called conditions of the EH boundary in \cite{AP2017,GBC}. 

From \r{WH} the dyadic $\=W$ must satisfy $\#m\.\=W=0$ and, hence, $\det\=W=0$. In the form \r{WH}, the general boundary conditions could be made unique by requiring an additional normalizing condition for the vector $\#m$. However, let us omit that for simplicity, whence the vector $\#m$ and the dyadic $\=W$ may be multiplied by an arbitrary scalar coefficient without changing the definition of the boundary.

The form \r{WH} resembles that of the impedance boundary \r{imp}, which can alternatively be written as \cite{Methods}
\e \#E_t = \=Z_t\.(\#n\x\#H_t), \f
or, as
\e \#n\x\#E = -(\=Z_t\xx\#n\#n)\.\#H, \f
in terms of what is normally called the impedance dyadic, $\=Z_t$. Because the vector $\#m$ is more general than the unit vector $\#n$, the form \r{WH} can be conceived as a generalization of the impedance-boundary conditions. 

The dyadic $\=W$ can be decomposed in its symmetric and antisymmetric parts,
\e \=W = \=W_s + \=W_a, \f
satisfying
\e \=W{}_s^T = \=W_s,\ \ \ \=W{}_a^T = -\=W_a. \f
They are defined by
 \e \=W_s = \frac{1}{2}( \#a_1\#b_2+\#b_2\#a_1-\#a_2\#b_1- \#b_1\#a_2), \l{Ws}\f
 \e \=W_a= \frac{1}{2}( \#a_1\#b_2-\#b_2\#a_1-\#a_2\#b_1+ \#b_1\#a_2). \f
The antisymmetric part can be represented as \cite{Methods}
\e \=W_a = \#w\x\=I,\f
in terms of the vector
\e \#w = \frac{1}{2}(\#b_2\x\#a_1+\#a_2\x\#b_1). \l{w}\f

\section{Duality Transformation}

In its basic form, the duality transformation in electromagnetic theory makes use of the symmetry of the Maxwell equations. This allows interchanging electric and magnetic quantities, $\#E\ra\#H$, $\#H\ra-\#E$, while the total set of equations remains the same. In this form it was originally introduced by Heaviside \cite{Heaviside}. In its more complete form, it can be defined by \cite{Methods} 
\e \am \#E_d \\ \h_o\#H_d\a = \amm A & B\\ C & D\a \am \#E\\ \h_o\#H\a, \l{dual}\f
where the four transformation parameters are assumed to satisfy
\e AD-BC=1. \f
The inverse transformation has the form
\e \am \#E \\ \h_o\#H\a = \amm D& -B\\ -C & A\a\am \#E_d\\ \h_o\#H_d\a. \l{invdual}\f

In addition to electromagnetic fields, \r{dual} induces transformations to electromagnetic sources and parameters of electromagnetic media and boundaries. One can show that, requiring the simple-isotropic medium with parameters $\E_o,\M_o$ to be invariant in the duality transformation, the parameters $A \cdots D$ must be chosen as  \cite{GBC,234}
\e A=D=\cos\VF,\ \ \ \ B=-C=\sin\VF, \l{ADBC}\f
which leaves us with one transformation parameter $\VF$, only. In the form \r{dual} and \r{ADBC}, the transformation was introduced by Larmor as duality rotation \cite{Fushchich,Mignaco}. 

Applying \r{invdual} and \r{ADBC}, the GBC conditions \r{GBC} are transformed to
\e \amm \#a_1 & \#b_1\\ \#a_2 & \#b_2\a \.\amm \cos\VF & -\sin\VF\\ \sin\VF & \cos\VF\a \am\#E_d\\ \h_o\#H_d\a = \am 0\\0\a, \l{GBC2d} \f
which yields the set of dual boundary conditions,
\e \amm \#a_{1d} & \#b_{1d}\\ \#a_{2d} & \#b_{2d}\a \.\am \#E_d\\ \h_o\#H_d\a = \am 0\\ 0\a,\l{GDBd}\f
in terms of the dual set of vectors  
\ea \#a_{1d} &=& \#a_1\cos\VF + \#b_1\sin\VF \l{a1d}\\
    \#b_{1d} &=& -\#a_1\sin\VF + \#b_1\cos\VF \\
    \#a_{2d} &=& \#a_2\cos\VF +\#b_2\sin\VF \\
    \#b_{2d} &=& -\#a_2\sin\VF + \#b_2\cos\VF. \l{b2d}\fa

Applying the duality transformation to boundary conditions of the form \r{WH} yields
\e \#m_d\x\#E_d = \=W_d\.\h_o\#H_d, \f
with
\ea \#m_d &=& \#a_{1d}\x\#a_{2d}\\
&=& (\#a_1\cos\VF + \#b_1\sin\VF)\x(\#a_2\cos\VF + \#b_2\sin\VF), \\
\=W_d &=& \#a_{1d}\#b_{2d}- \#a_{2d}\#b_{1d} \\
&=& (\#a_1\cos\VF + \#b_1\sin\VF)(\#b_2\cos\VF -\#a_2\sin\VF) \nonumber\\
&&-(\#a_2\cos\VF + \#b_2\sin\VF)(\#b_1\cos\VF-\#a_1\sin\VF). \fa
The dyadic $\=W_d$ can be expanded as
\e \=W_d = \=W\cos^2\VF + \=W{}^T\sin^2\VF + (\#a_1\x\#a_2- \#b_1\x\#b_2)\x\=I\ \cos\VF\sin\VF. \f
Remarkably, the symmetric part of $\=W_d$ equals that of $\=W$ in \r{Ws},
\e \=W_{ds} = \=W_s , \l{WdsWs}\f
while the antisymmetric part of $\=W_d$ written as
\e \=W_{da} = \#w_d\x\=I, \f
is related to \r{w} by
\e \#w_d = \#w\cos2\VF+ \frac{1}{2}(\#a_1\x\#a_2- \#b_1\x\#b_2)\sin2\VF.\f

\section{Self-Dual Boundary Conditions}

A number of special GBC boundaries have been previously shown to be self dual, i.e., invariant in the duality transformation. Such a property has engineering interest, because objects with certain geometric symmetry and self-dual boundary conditions appear invisible for the monostatic radar \cite{AP09}. In fact, in such cases there is no back scattering from the object for any incident wave. In particular, the SHDB boundary and its special cases, the SH and DB boundaries, are known to be self dual \cite{GBC}. Also, two special cases of the PEMC boundary, with $M= \pm 1/j\h_o$, have been shown to be self dual \cite{GBC}. The task of this paper is to define the most general class of GBC boundaries whose conditions are invariant in the duality transformation. 

For a boundary defined by conditions of the form \r{WH} to be self dual, we must have the three conditions
\ea \#m_d&=&\A\#m, \l{mdAm}\\
\#w_d &=& \A\#w, \l{wdAw}\\
\=W_{ds} &=& \A\=W_s, \l{WsAWs} \fa
valid for some scalar $\A$. The conditions \r{mdAm}, \r{wdAw} and \r{WsAWs} can be respectively expanded as
\e \#a_1\x\#a_2(\cos^2\VF-\A) + \#b_1\x\#b_2\sin^2\VF + (\#a_1\x\#b_2-\#a_2\x\#b_1)\sin\VF\cos\VF =0. \l{1}\f
\e (\#a_1\x\#b_2-\#a_2\x\#b_1)(\cos2\VF -\A) + (\#b_1\x\#b_2-\#a_1\x\#a_2)\sin2\VF=0.\l{2} \f
\e (1-\A)(\#b_2\#a_1 + \#a_1\#b_2 -\#b_1\#a_2 - \#a_2\#b_1)=0. \l{3}\f
Because from \r{WdsWs} and \r{WsAWs} we have
\e (\A-1)\=W_s=0, \f
for the boundary to be self dual, either $\A=1$ or $\=W_s=0$ must be satisfied. Let us consider these two cases separately.

\subsection{Case 1, $\A=1$}

The condition \r{3} is now satisfied identically, while the conditions \r{1} and \r{2} become
 \e (\#a_1\x\#a_2 - \#b_1\x\#b_2)\sin\VF + (\#b_2\x\#a_1+\#a_2\x\#b_1)\cos\VF =0, \l{1'}\f
\e -(\#b_2\x\#a_1 +\#a_2\x\#b_1)\sin\VF + (\#a_1\x\#a_2- \#b_1\x\#b_2)\cos\VF=0,\l{2'} \f
when excluding the identity transformation $\sin\VF=0$. After successive elimination of the bracketed terms and a comparison with \r{m} and \r{w}, the conditions \r{1'} and \r{2'} yield
\ea \#b_1\x\#b_2 = \#a_1\x\#a_2&=& \#m, \l{b1b2m}\\
\#b_2\x\#a_1 +\#a_2\x\#b_1&=& 2\#w=0, \l{b2a1}\fa
Because of  $\#w=0$, the case $\A=1$ requires that the dyadic $\=W$ be symmetric. 

From \r{b1b2m} and \r{b2a1}, it follows that the four vectors $\#b_1,\#b_2,\#a_1,\#a_2$ must be coplanar. Assuming $\#a_1\x\#b_1\not=0$, we can expand
\e \am \#a_2\\ \#b_2\a = \amm A_1 & B_1\\ A_2 & B_2\a \am \#a_1 \\ \#b_1\a . \l{A1B1A2B2}\f
Substituting these in  \r{b1b2m} and \r{b2a1}, we obtain the relations
\e A_2=-B_1,\ \ \ B_2=A_1. \f
The expansion \r{A1B1A2B2} now becomes
\e \am \#a_2\\ \#b_2\a = \amm A_1 & B_1\\ -B_1 & A_1\a \am \#a_1 \\ \#b_1\a , \f
whence we can write
\ea \#m &=& \#a_1\x\#a_2 = B_1\#a_1\x\#b_1,\\
\=W &=& \#a_1\#b_2-\#a_2\#b_1 = -B_1(\#a_1\#a_1+ \#b_1\#b_1), \fa
the latter of which has a symmetric form, as required. The boundary condition \r{WH} now becomes
\e (\#a_1\x \#b_1)\x\#E +(\#a_1\#a_1+ \#b_1\#b_1)\.\h_o\#H=0.\f
Denoting for simplicity $\#a_1$ and $\#b_1$, respectively, by $\#a$ and $\#b$, this is equivalent to the special form of the GBC conditions \r{GBC},
\e \amm \#a & \#b\\ -\#b & \#a\a \am \#E\\ \h_o\#H\a = \am 0\\ 0\a. \l{abcond}\f

To check the self-dual property of \r{abcond}, let us expand the dual boundary conditions \r{GDBd} in terms of \r{a1d} -- \r{b2d} as
$$ \amm \#a\cos\VF + \#b\sin\VF &  -\#a\sin\VF + \#b\cos\VF \\
    -\#b\cos\VF +\#a\sin\VF &  \#b\sin\VF + \#a\cos\VF\a \am \#E_d\\ \h_o\#H_d\a = $$
    \e \amm \cos\VF & -\sin\VF\\ \sin\VF & \cos\VF\a \amm \#a & \#b\\ -\#b & \#a\a \am \#E_d\\ \h_o\#H_d\a =\am 0\\ 0\a. \f
Obviously, these are equivalent to the original boundary conditions \r{abcond} whence the corresponding boundary is self dual for any parameter $\VF$.

\subsection{Case 2, $\=W_s=0$}

For $\A\not=1$, the dyadic $\=W$ must be antisymmetric and it can be expressed as
\e \=W = \#w\x\=I. \f
Applying \r{m} and \r{W} we have
\e \#m\.\=W=0,\ \ \Ra\ \ \#m\x\#w=0, \f
whence the two vectors must be linearly dependent,
\e \#w = \B\#m. \f
In this case, the boundary condition \r{WH} becomes that of the generalized PEMC boundary \cite{GPEMC},
\e \#m\x(\#H+ M\#E)=0, \l{GPEMC}\f
with $M=-1/\B\h_o$. For $\#m=\#n$, \r{GPEMC} is equivalent with the PEMC boundary condition, \r{PEMC}.

The dual boundary condition can be expanded as
\e\#m_d\x(\#E_d-\B\h_o\#H_d) = \A\#m\x((\cos\VF+\B\sin\VF)\#E + (\sin\VF-\B\cos\VF)\h_o\#H)=0. \f
To be self-dual, this should be a multiple of  \\
\e\#m\x(\#E-\B\h_o\#H), \f
which requires
\e \sin\VF-\B\cos\VF = -\B(\cos\VF+\B\sin\VF),\ \ \ \Ra\ \ \ \B^2=-1. \f
This leaves us with the two possibilities,
\e \B=\pm j. \f
In conclusion, in the case of antisymmetric dyadic $\=W$, the self-dual boundary condition must equal either of the two conditions 
\e \#m\x(\#E \pm j\h_o\#H)=0, \l{GPEMC1}\f
which can be called self-dual generalized PEMC boundaries with $M=\mp j/\h_o$.

To check the self-dual property of this result, we can apply \r{invdual} to \r{GPEMC1}. In fact, the resulting condition
\e e^{\pm j\VF}\#m\x(\#E_d \pm j\h_o\#H_d)=0, \l{GPEMCd}\f
equals that of \r{GPEMC1} for dual fields.

\subsection{Case 3, $\#m=0$}

The representation \r{WH} is not valid for $\#m=\#a_1\x\#a_2=0$, in which case the boundary conditions are of the reduced form \r{m=0}. In such a case, (see the Appendix) the self-dual condition requires that the vectors $\#b$ and $\#b_1$ be multiples of the vector $\#a$, whence the boundary conditions are reduced to the form
\e \amm \#a & 0\\ 0 & \#a\a \.\am \#E\\ \h_o\#H\a = \am 0\\ 0\a. \l{SDEH} \f
A boundary defined by conditions \r{SDEH} can be called a self-dual EH boundary because \r{SDEH} is the self-dual special case of \r{EH}, the EH-boundary conditions. Because \r{SDEH} is also a special case of \r{abcond}, we can actually include Case 3 as a subcase in the broader class of Case 1 boundaries.

\section{Special Cases}

Let us consider some special cases of the self-dual boundary of Case 1, as defined by the conditions \r{abcond}.

\begin{itemize}
\item $\#a=a\#n,\ \#b= b\#n$

In this case, the conditions \r{abcond} are reduced to those of the DB boundary, \r{DB}.
\item $\#a=\#b=\#a_t$

This case corresponds to that of the SH boundary, \r{SH}.
\item $\#a=\#a_t$, $\#b=\#b_t$, with $\#a_t\x\#b_t=\#n$

The conditions \r{abcond} can now be written as 
\e \#n\x\#E = -\h_o(\#a_t\#a_t+ \#b_t\#b_t)\.\#H, \f
which has the form of impedance-boundary conditions, \r{imp}. The impedance dyadic is symmetric,
\e \=Z_t = \h_o\#n\#n\xx(\#a_t\#a_t+ \#b_t\#b_t),\f
and satisfies \cite{GBC}
\e \det_t\=Z_t = \tr\=Z{}_t^{(2)} = \h_o^2. \f
The impedance dyadic consists of isotropic and anisotropic parts as \cite{GBC}
\e \=Z_t=Z_s\=I_t + \=Z_a, \f
with
\e \=I_t=\=I -\#n\#n, \f
and
\ea Z_s &=& \frac{1}{2}\tr\=Z_t= \frac{\h_o}{2}(\#a_t\.\#a_t + \#b_t\.\#b_t), \\
 \=Z_a &=& \=Z_t - Z_s\=I_t. \fa
An example of this type of self-dual surface is the perfect co-circular polarization reflector \cite{Jensen}.

\item $\#a=a_n\#n,\#b=\#b_t$

This case corresponds to that of the SHDB boundary, \r{SHDB}. A similar case is obtained for $\#a=\#a_t,\#b=b_n\#n$.

\item $\#a=a_n\#n + \#a_t$, $\#b= \#b_t$

Actually, the case that one of the vectors $\#a$ and $\#b$ has only the tangential component, in general equals the case \r{abcond}. In fact, subsequently eliminating $\#n\.\#E$ and $\#n\.\#H$ from \r{abcond}, the remaining equations, with redefined vectors $\#a$ and $\#b$, can be expressed in the form
\e \amm \#a & \#b_t\\ -\#b_t& \#a\a \.\am \#E\\ \h_o\#H\a=\am 0\\ 0\a, \l{SDBC}\f
where $\#b$ has no normal component. 

\item $\#a=\#b$

This case corresponds to that of the self-dual EH boundary, defined by conditions of the form \r{SDEH}.

\end{itemize}

\section{Plane-Wave Reflection from Self-Dual Boundary}

Let us consider a plane wave incident to and reflecting from a boundary surface defined by the self-dual conditions \r{abcond}. The respective $\#k$ vectors 
\ea \#k^i &=& -\#n k_n + \#k_t,\\
\#k^r &=& \#n k_n + \#k_t, \fa
satisfy
\e \#n\.\#k_t=0,\ \ \ \#k^i\.\#k^i=\#k^r\.\#k^r=k_o^2. \f
Following the analysis of \cite{GBC}, Sec.\ 5.4,  we can express the reflected electric field in terms of the incident electric field as
\e \#E^r = \=R\.\#E^i, \l{ERE}\f
where the reflection dyadic has the form
\e \=R = \frac{1}{J^r}\#k^r\x(\#c_2^r\#c_1^i-\#c_1^r\#c_2^i), \l{R} \f
with
\ea \#c_1^{i,r} &=& \#k^{i,r}\x\#b_1 -k_o\#a_1= \#k^{i,r}\x\#b -k_o\#a, \\
\#c_2^{i,r} &=& \#k^{i,r}\x\#b_2 -k_o\#a_2=  \#k^{i,r}\x\#a +k_o\#b, \fa
and
\e J^r = \#k^r\.\#c_1^r\x\#c_2^r. \l{Jr}\f

Now it is quite straightforward to show that, if the incident field is decomposed in two parts as
\e \#E^i = \#E_1^i+ \#E_2^i, \l{Ei12}\f
and defined by
\e \#E_1^i = E_1^i\#k^i\x\#c_1^i,\ \ \ \#E_2^i = E_2^i \#k^i\x\#c_2^i,\f
from \r{ERE} and \r{R}, the reflected field will be decomposed as
\e \#E^r = \#E_1^r + \#E_2^r, \f 
and defined by
\e \#E_1^r = E_1^r\#k^r\x\#c_1^r,\ \ \ \#E_2^r = E_2^r \#k^r\x\#c_2^r.\f
The four field vectors satisfy $\#c_1^i\.\#E_1^i=\#c_2^i\.\#E_2^i=0$ and $\#c_1^r\.\#E_1^r=\#c_2^r\.\#E_2^r=0$, while the scalar coefficients are obtained from the field vectors as
\e E_1^i= \frac{1}{J^i}\#c_2^i\.\#E^i,\ \ \ \ E_2^i = -\frac{1}{J^i}\#c_1^i\.\#E^i, \f
\e E_1^r= \frac{1}{J^r}\#c_2^r\.\#E^r,\ \ \ \ E_2^r = -\frac{1}{J^r}\#c_1^r\.\#E^r, \f
with
\e J^i = \#k^i\.\#c_1^i\x\#c_2^i. \f

Substituting  \r{Ei12} and \r{R} in \r{ERE}, relations between the scalar field coefficients can be written as
\ea J^rE_1^r &=&  -J^iE_1^i \l{E1rE1i}\\
J^rE_2^r &=& -J^iE_2^i.\l{E2rE2i} \fa
Thus, there is no cross coupling between the waves 1 and 2 in reflection from the boundary, and the ratio of two scalar field coefficients is the same,
\e \frac{E_1^r}{E_1^i} = \frac{E_2^r}{E_2^i}= -\frac{J^i}{J^r}. \l{ref}\f 
Here we have assumed $J^r\not=0$ and $J^i\not=0$. 

In the case $J^i=0$, i.e., if the wave vector $\#k=\#k^i$ satisfies
\e J(\#k)= (\#k\x\#a)^2 + (\#k\x\#b)^2 + 2k_o\#k\.\#a\x\#b=0, \l{Jk} \f
from \r{E1rE1i} and \r{E2rE2i} it follows that $E_1^r=E_2^r=0$, that is, $\#E^r=0$. Thus, for such a wave vector, the incident wave $\#E^i,\#H^i$ satisfies the boundary conditions identically, i.e., it is matched to the boundary. Similarly, for $J^r=0$, when the wave vector $\#k^r$ satisfies $J^r=0$, which equals \r{Jk} for $\#k=\#k^r$, the incident wave vanishes, $\#E^i=0$. In fact, from \r{R} it follows that, for $J^r=0$, the magnitude of the reflection dydic becomes infinite, whence for finite $\#E^r$, we must have $\#E^i=0$. In this case the reflected wave $\#E^r,\#H^r$, is matched to the boundary. For any single matched plane wave it does not matter whether it is called "incident" or "reflected". Thus, the reflection coefficient \r{ref} may be either zero or infinite for the matched-wave cases. \r{Jk} is called the dispersion equation for the matched waves of the boundary\cite{GBC}.

\section{Matched Waves for Self-Dual EH Boundary}

As a more concrete example, let us consider the self-dual EH boundary defined by \r{SDEH}, i.e., by \r{abcond} with $\#b=0$. In this case, we can write
\ea \#c_1^{i,r} &=& -k_o\#a, \\
\#c_2^{i,r} &=&  \#k^{i,r}\x\#a, \fa
whence, from \r{Jr}, $J^r=-k_o(\#k^r\x\#a)^2$. The reflection dyadic \r{R} can be represented in the form
\e \=R = \frac{1}{(\#k^r\x\#a)^2}((\#k^r\x(\#k^r\x\#a))\#a -(\#k^r\x\#a)(\#k^i\x\#a)). \l{REH}\f

Let us now assume that $\#a=\#u$ is a real unit vector, and $(\#u,\#v,\#w)$ is a right-hand basis of real orthogonal unit vectors. Denoting 
\e \#k = k_u\#u + \#k_\bot,\ \ \ \ \#k_\bot\.\#u=0, \f
the dispersion equation \r{Jk} now becomes
\e (\#k\x\#a)^2 = ((k_u\#u+ \#k_\bot)\x\#u)^2 =\#k_\bot\.\#k_\bot=0, \f
whence $\#k_\bot$ may be any circularly-polarized vector in the plane orthogonal to $\#u$. From
\e \#k\.\#k = k_u^2+ \#k_\bot\.\#k_\bot = k_u^2= k_o^2, \f
we obtain $k_u=k_o$ or $k_u=-k_o$.

Now any circularly-polarized vector orthogonal to $\#u$ can be represented as a multiple of one of the two circularly polarized vectors $\#u_+,\#u_-$ defined by \cite{Methods}
\e \#u_+ = \#v+ j\#w,\ \ \ \ \#u\x\#u_+ = -j\#u_+,\f
\e \#u_-= \#v- j\#w,\ \ \ \ \#u\x\#u_- = j\#u_-. \f
as
\e \#k_+{}_\bot = k_+\#u_+,\ \ \ \ \#k_-{}_\bot = k_-\#u_-. \f
Assuming $\#n\.\#u>0$, we can express the possible vectors $\#k^i$ satisfying $J^i=0$ by
\e \#k^i_+= -k_o\#u + k_+^i\#u_+,\ \ \ \ \#k^i_-= -k_o\#u + k_-^i\#u_-. \l{kipm} \f
Similarly, the possible vectors $\#k^r$ satisfying $J^r=0$ can be expressed as
\e \#k^r_+= k_o\#u + k_+^r\#u_+,\ \ \ \ \#k^r_-= k_o\#u + k_-^r\#u_-. \l{krpm}\f
We can now make use of the following relations between the fields at the boundary,
\ea \#E^r &=& \frac{1}{(\#k^r\x\#a)^2}\#k^r\x(\#k^r\x\#a\#a+\#a\#a\x\#k^i)\.\#E^i, \l{Er}\\
    \#E^i &=& \frac{1}{(\#k^i\x\#a)^2}\#k^i\x(\#k^i\x\#a\#a+\#a\#a\x\#k^r)\.\#E^r. \l{Ei}\fa
The relation \r{Er} is obtained from \r{REH}, while \r{Ei} can be verified by eliminating $\#E^i$ from the two equations, which yields $\#E^r=\#E^r$. 

The fields for the matched waves at the self-dual EH boundary can be found for the two cases from \r{Er} and \r{Ei} as follows:

\begin{enumerate}
\item For $J^i=0$ and $\#k^i=\#k^i_\pm= -k_o\#u + k_\pm^i\#u_\pm$, from \r{Er} we obtain $\#E^r_\pm=0$ for the incident fields 
\ea \#E^i_\pm &=& \A_\pm^i \#u\x(\#u\x\#k^i_\pm) = E^i_\pm\#u_\pm \\
\h_o\#H_\pm^i &=& \frac{1}{k_o}\#k_\pm^i\x\#E_\pm^i = -\#u\x E_\pm^i\#u_\pm = \pm j\#E_\pm^i. \fa
\item For $J^r=0$ and $\#k^r=\#k^r_\pm= k_o\#u + k_\pm^r\#u_\pm$, from \r{Ei} we obtain $\#E^i_\pm=0$ for the reflected fields
\ea \#E^r_\pm &=& \A_\pm^r \#u\x(\#u\x\#k^r_\pm) = E^r_\pm \#u_\pm\\
\h_o\#H_\pm^r &=& \frac{1}{k_o}\#k_\pm^r\x\#E_\pm^r = \#u\x \#E^r_\pm =\mp j\#E_\pm^r.   \fa
This "reflected" matched wave corresponds to the non-existing "incident" wave arriving at $\#k^i= (\=I-2\#n\#n)\.\#k^r$.
\end{enumerate}
In conclusion, the plane waves matched to the self-dual EH boundary are circularly polarized parallel to the plane orthogonal to $\#a=\#u$.

For $\#a=\#u=\#n$, the self-dual EH boundary is reduced to the DB boundary. It is known that the normally incident or reflected wave is matched to the DB boundary for any polarization \cite{GBC}.

\section{Reflection from Self-Dual EH Boundary}

Let us consider reflection from the self-dual EH boundary, defined by the conditions \r{SDEH}, applying the relation \r{Er}. Such a boundary has previously found research interest and its realization in terms of a medium interface has been suggested \cite{Tedeschi}. $\#a$ is assumed to be a real unit vector and form the angle $\A$ with the normal $\#n$ of the boundary, Figure~\ref{fig:geometria},
\e \#u = \#n\cos\A + \#u_1\sin\A. \f
Here, $\#u_1$ denotes a real unit vector tangential to the boundary and $\#u_2=\#n\x\#u_1$. 
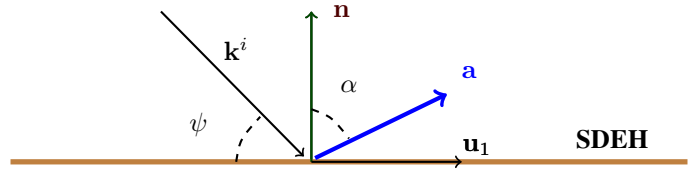
\begin{figure}[htbp]
                             \centering
                             \begin{tikzpicture}[scale=1.0]
                             \draw[thick][->] (0,0) -- (0,2);
                             \node at (2.2,.2) {$\mathbf{u}_1$};
                             \draw[brown, line width=2pt] (-4,0) -- (5,0);
                             \node[black] at (4,.3) {\bf{SDEH}};
                             \draw[thick,green!30!black] [->] (0,0)--(0,2);
                             \node at (-1,1.5) {$\mathbf{k}^i$};
                             \node[blue] at (2.1,1.2) {$\mathbf{a}$};
                             \node at (-1.5,.5) {$\psi$};
                             \node at (.5,1) {$\alpha$};
                             \node[red!30!black] at (.4,2) {$\mathbf{n}$};
                             \draw[thick] [->] (-2,2)--(-.1,5/70);
                             \draw[blue, ultra thick] [->] (.05,.05)--(1.8,.9);
                             \draw[thick,dashed] [-] (-1,-0) .. controls (-.95,.3) .. (-.68,0.6); 
                             \draw[thick,dashed] [-] (0,.7) .. controls (.27,.6) .. (.5,0.31); 
                             \draw[thick][->] (0,0) -- (2,0);
                        \node at (2.2,.2) {$\mathbf{u}_1$};
                             \end{tikzpicture}
                             \caption{Plane wave incident on a self-dual EH boundary. The incident wave forms the elevation angle $\psi$ with the surface. $\#a$ is a real unit vector.}
                             \label{fig:geometria}
\end{figure}

Let us separate two main cases of incidence to the self-dual EH boundary:  $\#k^i$ lies in a plane either parallel or perpendicular to the plane of $\#a$ and $\#n$.

\subsection{Parallel Incidence}

For the wave incident in the plane of $\#n$ and $\#a$ at an angle $\psi$,  Figure~\ref{fig:geometria}, we can write
\ea \#k^i &=& k_o(-\#n\sin\psi + \#u_1 \cos\psi) = k_o\#u_2\x\#u_p^i, \\
\#k^r &=& k_o(\#n\sin\psi + \#u_1 \cos\psi)= k_o\#u_2\x\#u_p^r, \fa
where the two unit vectors
\ea \#u_p^i &=& \#n\cos\psi + \#u_1\sin\psi, \\
  \#u_p^r &=& -\#n\cos\psi + \#u_1\sin\psi, \fa
satisfy 
\e  \#u_p^i\.\#k^i=\#u_p^i\.\#u_2=0,\ \ \ \#u_p^r\.\#k^r=\#u_p^r\.\#u_2=0, \f

Inserting these in \r{Er}, we obtain
\e \#E^r = -\frac{1}{\cos(\psi+\A)}\big(-\#u_p^r(\#n\cos\A +\#u_1\sin\A)+ \#u_2\#u_2\cos(\psi-\A)\big)\.\#E^i. \l{ErEi}\f 
Expanding the incident and reflected fields as
 \ea \#E^i&=& \#u_p^iE^i_p + \#u_2E^i_2,\l{Eip2} \\
 \#E^r&=& \#u_p^rE^r_p + \#u_2E^r_2, \fa
in their components $E_p^i,E_p^r$, and  $E^i_2,E^r_2$, respectively parallel and perpendicular to the plane of $\#a$ and $\#k^i$, the expression for the reflected field \r{ErEi} can be given the simple form
 \e \#E^r = \frac{\cos(\psi-\A)}{\cos(\psi+\A)}(\#u_p^r E_p^i - \#u_2E_p^i)). \f
Because the field magnitudes obey the relations
\e E_p^r =  R_p E_p^i,\ \ \ \ E_2^r =  R_2 E_2^i,\ \ \ \ R_p=-R_2=\frac{\cos(\psi-\A)}{\cos(\psi+\A)}, \l{129}\f
the reflection coefficient has the same magnitude for the two polarizations. 

For $\#a=\#n$ we have $\A=0$, in which case the self-dual EH boundary equals the DB boundary. In this case, the parallel polarization is reflected as from the PMC boundary ($R_p=1$), and the perpendicular polarization as from the PEC boundary ($R_2=-1$). 

For $\psi=\pi/2+\A$ we have $\#E^r=0$ and, for $\psi=\pi/2-\A$ we have $\#E^i=0$, which correspond to the respective two cases of matched waves $J^i=0$ and $J^r=0$.

\subsection{Perpendicular Incidence}

For the wave incident with $\#k^i$ in the plane normal to the vector $\#u_1$, we have
\ea \#k^i &=& -\#n\sin\psi + \#u_2 \cos\psi, \\
\#k^r &=& \#n\sin\psi + \#u_2 \cos\psi. \fa
Because the wave vectors satisfy the property
\e (\#k^i\x\#a)^2 = (\#k^r\x\#a)^2 = k_o^2(\cos^2\psi + \sin^2\psi\sin^2\A), \l{kikru}\f
there are no matched waves for real $\psi$ values if $\A$ has a real value. 

Let us expand the two fields as
\ea \#E^i &=& (\#k^i\x\#a)A^i + \#k^i\x(\#k^i\x\#a)B^i,\l{EiAi} \\
\#E^r &=& (\#k^r\x\#a)A^r + \#k^r\x(\#k^r\x\#a)B^r. \l{ErAr}\fa
The coefficients can be found from
\ea A^{i,r} &=& \frac{\#k^{i,r}\x\#a}{(\#k^{i,r}\x\#a)^2}\.\#E^{i,r} = -\frac{k_o\#a\.\h_o\#H^{i,r}}{(\#k^{i,r}\x\#a)^2}, \\
 B^{i,r} &=& \frac{\#k^{i,r}\x(\#k^{i,r}\x\#a)}{k_o^2(\#k^{i,r}\x\#a)^2}\.\#E^{i,r} = -\frac{\#a\.\#E^{i,r}}{(\#k^{i,r}\x\#a)^2}, \fa

Substituting \r{EiAi} in \r{Er} and taking \r{kikru} into account, we can expand
\ea \#E^r &=& \frac{1}{(\#k^r\x\#a)^2}\#k^r\x(\#k^r\x\#a\#a-\#a\#k^i\x\#a)\.\#E^i \nonumber\\
&=&\frac{1}{(\#k^i\x\#a)^2}(\#k^r\x(\#k^r\x\#a)(-(\#k^i\x\#a)^2B^i)-(\#k^r\x\#a)(\#k^i\x\#a)^2A^i) \nonumber\\
&=& -(\#k^r\x\#a)A^i-\#k^r\x(\#k^r\x\#a)B^i. \l{Erexp} \fa
Comparing with the expansion \r{ErAr}, we finally obtain the simple relations between the field coefficients,
\e A^r=-A^i,\ \ \ \ B^r=-B^i. \l{ArAiBrBi}\f
These relations are independent of the polarization of the incident field $\#E^i$. Because we can write
\e \#E^r\.\#E^r = (\#k^r\x\#a)^2A^{i2} + k_o^2(\#k^r\x\#a)^2 B^{i2} =\#E^i\.\#E^i, \f
for real $\#k^i$, the magnitude of the field in reflection is constant for all angles of incidence. This effect is demonstrated in Figure \ref{fig:numerics}.

\subsection{Numerical Example}
 
As an example of the characteristics of a self-dual EH boundary, in the following, reflection amplitudes of plane waves are computed for different incidence angles and polarizations. The vector $\#a$ defining the boundary according to \r{SDEH} forms the angle $\A=\pi/3=60^\circ$ with the normal $\#n$, as shown in Figure~\ref{fig:geometria}.
Figure~\ref{fig:numerics} displays the magnitude of the reflection coefficient as function of the elevation angle $\psi$, defined in Figure~\ref{fig:geometria}. The magnitude of the reflection coefficient is computed for two planes of incidence: the plane spanned by $\mathbf n$ and $\mathbf a$ (the plane in Figure~\ref{fig:geometria}, with solid blue line), and the plane perpendicular to that (plane of $\#n$ and $\#n\x\#a$, with dashed orange line).

\begin{figure}
	\centerline{\includegraphics[width=5cm]{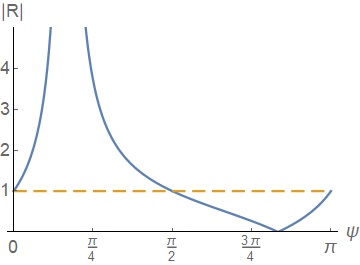}
	\hspace{10pt}\includegraphics[width=6cm]{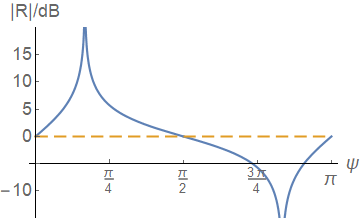}}
	\caption{The magnitude of the reflection coefficient 
	$|R|$ (absolute and logarithmic) for a plane wave incident to a planar self-dual EH boundary
 defined by a real vector $\mathbf a$ which forms the angle $\pi/3$ with the normal of the boundary. 
 Solid blue line: $\mathbf k^i$ lies in the plane parallel to the vectors $\mathbf n$ and $\mathbf a$ and makes the angle $\psi$ with the boundary (see Figure~\ref{fig:geometria}). Matched wave conditions appear for $\psi=5\pi/6$ (matched incident wave) and for $\psi=\pi/6$ (matched reflected wave). Dashed orange line: the  reflection coefficient is $|R|=1$ for all incidence angles in the plane parallel to vectors 
 $\mathbf n$ and $\mathbf n \times\mathbf a$. The value $|R|$ is independent of the polarization of the wave in both cases.
}
	\label{fig:numerics}
\end{figure}

The results show the extremely strong dependence of the response on the direction of the incident wave. While in the plane perpendicular to that of Figure~\ref{fig:geometria} the reflection satisfies $|R|=1$ (meaning that all power is reflected), the angular dependence in the $\mathbf n$-$\mathbf a$ plane contains a zero and an infinity. These are the two matched-wave cases. For the elevation angle $\psi= \pi/2+\alpha=5\pi/6$, the plane wave is incident from the direction of $\mathbf a$, whence $\#a\.\#E=\#a\.\#H=0$. Because the boundary conditions are satisfied, the wave is matched, and no reflection is generated. Likewise, for the case $\psi=\pi/2-\alpha=\pi/6$, the reflected wave is matched, leading to a singularity in the reflection coefficient.

As shown in Equations~\r{129} and \r{ArAiBrBi}, the magnitude of the reflection coefficients is independent of the polarization of the incident wave. For the dashed orange line with full reflection independently of elevation angle, the reflection coefficient is $R=-1$ for a perpendicularly (TE) polarized wave while for the parallel-polarized (TM) incidence, the reflection coefficient is $R=+1$. This behavior (PEC-like for TE-polarization and PMC-like for TM-polarization) is exactly the same as for a DB boundary \cite{DB,259}. It is worth noting that these two polarizations are the eigenpolarizations in reflection: for an arbitrary incident wave, the polarization changes in general but nevertheless the reflection amplitude remains at $|R|=1$.

Also in the other plane (plane of Figure~\ref{fig:geometria}, solid blue curve), the reflection magnitude is polarization-independent. In fact, also for any incidence direction, not necessarily in these two planes, the same observation applies that the magnitude of the reflection coefficient is independent of the polarization state of the wave.

\section{Conclusion}

It has been shown in this study that the possible self-dual electromagnetic boundaries, invariant in duality transformations,  fall in two possible classes: Case 1, those which are certain generalizations of the soft-and-hard/DB (SHDB) boundaries, as defined by conditions of the form \r{abcond}, and Case 2, those called self-dual generalized PEMC (GPEMC) boundaries, defined by conditions of the form \r{GPEMC}. Plane-wave reflection and matched waves associated to these boundaries have been analyzed and numerical examples were computed for the self-dual EH boundary, which belongs to the class of Case 1 boundaries. 

\section*{Appendix: Self-Dual EH Boundary}

Let us consider in more detail the condition 
\e \#m = \#a_1\x\#a_2 =0, \f
making the representation  \r{WH} invalid and find the corresponding self-dual boundary conditions. Here we can separate the three cases:

\begin{itemize}
\item $\#a_1=\#a_2=0$.

In this case, \r{GBC} yields
\e \#b_1\.\#H=\#b_2\.\#H=0, \f
i.e., conditions of the H-boundary \cite{GBC}, which are not self dual.
\item $\#a_1\not=0$ and $\#a_2=0$, (the converse case can be handled similarly). 

The boundary conditions \r{GBC} can be written in the form
\ea \#a\.\#E + \#b_1\.\h_o\#H &=& 0, \l{b1H}\\
\#b\.\h_o\#H &=& 0. \l{bH}\fa
\item $\A_1\#a_1+\A_2\#a_2=0$ with $\A_1\not=0,\ \A_2\not=0$.

In this case, after elimination, \r{GBC} can be reduced to the previous form \r{b1H} and \r{bH}.
\end{itemize}

The self-dual conditions for a boundary defined by \r{b1H} and \r{bH}, can be found by requiring that the dual set of vectors \r{a1d} - \r{b2d} satisfy 
\e \amm \#a\cos\VF + \#b_1\sin\VF & -\#a\sin\VF + \#b_1\cos\VF\\ \#b\sin\VF & \#b\cos\VF\a = \amm \A & \B\\ \g & \de\a \amm \#a & \#b_1\\ 0 & \#b \a, \f 
for some scalars $\A\cdots\de$. These can be written as
\ea \#a\cos\VF + \#b_1\sin\VF &=& \A\#a,\l{ab1} \\
\#b\sin\VF &=& \g\#a, \l{bsga}\\
-\#a\sin\VF + \#b_1\cos\VF &=& \A\#b_1 + \B\#b, \\
\#b\cos\VF &=& \g\#b_1 + \de\#b. \fa
Since $\sin\VF=0$ corresponds to identity transformation, and $\#b=0$ to incomplete boundary conditions, these possibilities can be neglected. From \r{bsga} and \r{ab1} we have, respectively, 
\ea \#b&=&\frac{\g}{\sin\VF}\#a. \l{bga}\\
 \#b_1 &=& \frac{\A-\cos\VF}{\sin\VF}\#a. \fa
Since $\#b$ is a multiple of $\#a$, and $\#b_1$ is a multiple of $\#a$ or zero, in the self-dual case, the boundary conditions \r{b1H}, \r{bH} are reduced to the form \r{SDEH}, corresponding to those of the self-dual EH boundary.


\begin{thebibliography}{99}

\bibitem{PIERL2016} I.V. Lindell and A. Sihvola, "Electromagnetic boundaries with PEC/PMC equivalence," {\it Prog. Electromag. Res. Lett.}, vol. 61, pp. 119-123, 2016.

\bibitem{AP2017}  I.V. Lindell and A. Sihvola, "Electromagnetic wave reflection from boundaries defined by general linear and local conditions," {\it IEEE Trans.\ Antennas Propag.}, vol. 5, no. 9 pp. 4656-4663, September 2017.

\bibitem{GBC} I.V. Lindell and A. Sihvola, {\it Boundary Conditions in Electromagnetics}, Hoboken, NJ: Wiley and IEEE Press, 2020.

\bibitem{PEMC} I.V. Lindell and A. Sihvola, "Perfect electromagnetic conductor", {\it J. Electro. Waves Appl.}, Vol.19, No.7, pp.861--869, 2005.

\bibitem{253} A. Sihvola and I. V. Lindell, "Perfect electromagnetic conductor medium." {\it Ann. Phys.} (Berlin) Vol.17, pp.787--802, September/October 2008.

\bibitem{Methods} I.V. Lindell, {\it Methods for Electromagnetic Field Analysis}, 2nd ed., Piscataway, N.J.:  Wiley and IEEE Press, 1995.

\bibitem{SH} P.-S. Kildal, "Definition of artificially soft and hard surfaces in electromagnetics," {\it Electron. Lett.}, Vol. 24, pp. 168--170, 1988.

\bibitem{SH1} P.-S. Kildal, "Artificially soft and hard surfaces in electromagnetics", {\it IEEE Trans. Antennas Propagat.}, Vol. 38, No. 10, pp. 1537--1544, Oct. 1990.

\bibitem{GSH} I.V. Lindell, "Generalized soft-and-hard surface," {\it IEEE Trans. Antennas Propag.}, vol. 50, No. 7, pp. 926--929, 2002.

\bibitem{Rumsey} V.H. Rumsey, "Some new forms of Huygens' principle", {\it IRE Trans. Antennas Propag.}, Vol. 7, pp. S103--S116, Dec. 1959.

\bibitem{DB} I.V. Lindell and A. Sihvola, "DB boundary as isotropic soft surface," {it Proc. Asian Pacific Microwave Conf.}, Hong Kong, Dec. 2008 (4 pages).


\bibitem{259} I.V. Lindell and A. Sihvola, "Electromagnetic boundary conditions defined in terms of normal field components," {\it IEEE Trans.\ Antennas Propag.}, Vol.58, no.4, pp.1128--1135, April 2010. 

\bibitem{SHDB} I.V. Lindell and A. Sihvola, "Soft-and-hard/DB boundary conditions realized by a skewon-axion medium," {\it IEEE Trans. Antennas Propag.}, Vol. 61, no. 2, pp. 768--774, Feb. 2013.

\bibitem{GSHDB} I.V. Lindell and A. Sihvola, "Generalized Soft-and-Hard/DB Boundary," {\it IEEE Trans.\ Antennas Propag.}, vol. 65, no.1 pp. 226-233, January 2017.

\bibitem{Heaviside} O. Heaviside, {\it Electrical Papers}, New York: Chelsea 1970; reprint of the first edition, London 1892 (Vol. 1, p. 447; Vol. 2, pp. 172--175). The original articles were respectively published in {\it The Electrician}, 1885, p. 306 and {\it Phil. Mag.}, Aug. 1886, p. 118.
 
\bibitem{234} I.V. Lindell and A.H. Sihvola, "Transformation method for problems involving perfect electromagnetic conductor (PEMC) structures," {\it IEEE Trans.\ Antennas Propag.}, Vol.53, no.9, pp.3005-3011, September 2005.

\bibitem{Fushchich} W.I. Fushchich and A.G. Nikitin, "On the new symmetries of Maxwell equations," {\it Czech. J. Phys.}, Vol B32, pp. 476--480, 1982.

\bibitem{Mignaco} J.A. Mignaco, "Electromagnetic duality, charges, monopoles, topology, ...", {\it Braz. J. Phys.}, Vol. 31, no.2, 2001.

\bibitem{AP09} I.V. Lindell, A. Sihvola, P. Yl\"a-Oijala and H. Wallen "Zero backscattering from self-dual objects of finite size," {\it IEEE Trans.\ Antennas Propag.}, Vol. 57, No.9, pp.2725--2731, Sept. 2009.


\bibitem{GPEMC} I.V. Lindell and A. Sihvola, "Generalization of Perfect Electromagnetic Conductor (PEMC) Boundary," Submitted to {\it IEEE Trans.\ Antennas Propagat.}, December 2019.

\bibitem{Jensen} F. Liu, S. Xiao, A. Sihvola and J. Li, "Perfect co-circular polarization reflector: A class of reciprocal perfect conductors with total co-circular polarization reflection," {\it IEEE Trans. Antennas Propag.,} Vol. 62, no. 12, pp. 6274--6281, 2014.

\bibitem{Tedeschi} N. Tedeschi, F. Frezza, A. Sihvola: "Reflection and transmission at the interface with an electric–magnetic uniaxial medium with application to boundary conditions", {\it IEEE Trans. Antennas Propag.}, Vol. 61, No. 11, pp. 5666--5675, Nov. 2013.































\end{thebibliography}
\end{document}